# Hexagonal Boron Nitride-Graphene Heterostructures with Enhanced Interfacial Thermal Conductance for Thermal Management Applications

Saheb Karak[1], Suvodeep Paul[1], Devesh Negi[1], Bommareddy Poojitha[1], Saurabh Kumar Srivastav[2], Anindya Das[2], and Surajit Saha[1*]

[1]Department of Physics, Indian Institute of Science Education and Research, Bhopal, 462066, India

[2]Department of Physics, Indian Institute of Science, Bangalore, 560012, India

*Correspondence: surajit@iiserb.ac.in

***Abstract:*** Atomically thin monolayers of graphene show excellent electronic properties which have led to a great deal of research on their use in nanoscale devices. However, heat management of such nanoscale devices is essential in order to improve their performance. Graphene supported on hexagonal boron nitride (h-BN) substrate has been reported to show enhanced (opto)electronic and thermal properties as compared to extensively used $SiO_2$/Si supported graphene. Motivated by this, we have performed temperature- and power-dependent Raman Spectroscopic measurements on four different types of (hetero)structures: (a) h-BN (BN), (b) graphene (Gr), (c) h-BN on graphene (BG), and (d) graphene encapsulated by h-BN layers from both top and bottom (BGB), all supported on $SiO_2$/Si substrate. We have estimated the values of thermal conductivity ($\kappa$) and interfacial thermal conductance per unit area (g) of these four (hetero)structures to demonstrate the structure-activity (thermal) relationship. We report here the values of $\kappa$ and g for h-BN supported on $SiO_2$/Si as 280.0±58.0 $Wm^{-1}K^{-1}$ and 25.6±0.4 $MWm^{-2}K^{-1}$, respectively. More importantly, we have observed an improvement in both thermal conductivity and interfacial thermal conductance per unit area in the heterostructures which ensures a better heat dissipation in devices. The $\kappa$ and g of h-BN encapsulated graphene on $SiO_2$/Si (BGB) sample was observed to be 850.0±81.0 $Wm^{-1}K^{-1}$ and 105±1 $MWm^{-2}K^{-1}$, respectively, as opposed to 600.0±93.0 $Wm^{-1}K^{-1}$ and 1.15±0.40 $MWm^{-2}K^{-1}$, respectively, for graphene on $SiO_2$/Si substrate. Therefore, we propose that for graphene-based nanoscale devices, encapsulation with h-BN is a better alternative to address heat management issues.

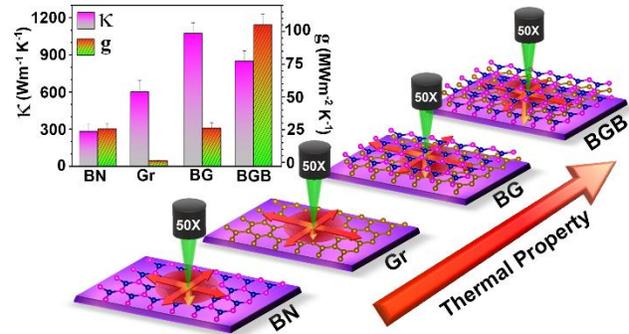

***Keywords:*** Nanoscale, Raman spectroscopy, interfacial thermal conductance per unit area, thermal conductivity, encapsulation, h-BN/graphene heterostructure.

*Introduction:*

Since the discovery of graphene in 2004 [1], it has been extensively studied because besides being the first atomically thin two-dimensional (2D) material known to man, it showed intriguing properties like exceptionally high charge carrier mobility (250000 cm$^2$/V s) [2], high thermal conductivity (5000 W/m K) for suspended graphene [3], good electrical conductivity, flexibility, large surface to volume ratio, and high modulus of elasticity (~1 TPa) [4]. These extraordinary properties led to the use of graphene in various important potential applications including the miniaturization of electronic devices [1-9]. However, easy fabrication of such miniaturized nanoscale devices often requires graphene to be supported on a substrate (as compared to suspended devices) which deteriorates the electronic and thermal properties of graphene due to lattice mismatch and interaction with the substrate. Reports of graphene supported on the most commonly used SiO$_2$/Si substrate shows a degradation of both the charge carrier mobility and thermal conductivity of graphene [10]. Therefore, researchers are in a search of alternative substrates for graphene-based nanoscale devices.

Hexagonal boron nitride (h-BN) substrate (which shows excellent lattice matching with graphene) reportedly enhances the electron mobility of graphene up to 60000 cm$^2$/Vs [11], as opposed to 10000 cm$^2$/Vs [11] when supported on SiO$_2$/Si substrate. Further, h-BN has properties like, (a) inertness, (b) absence of dangling bonds and surface charge traps, (c) atomically planar surface, and (d) dielectric properties, etc. [11-16], thus proving its relevance as a substrate (or buffer layer) rather than SiO$_2$/Si for graphene-based devices. Besides, the enhanced electronic properties, good performance of nanoscale devices also demand the addressing of thermal issues which involves proper dissipation of heat. Efficient dissipation of heat in substrate supported graphene devices may be two-fold. The first channel of heat dissipation is through the graphene channel itself (which demands a high in-plane thermal conductivity, κ) and a cross-plane dissipation through the substrate (characterized by the interfacial thermal conductance per unit area, g). While the κ value for graphene degrades when supported on a substrate, the new channel of cross-plane heat dissipation through the substrate is an extra benefit. Therefore, the requirement is the choice of a substrate that will provide higher cross-plane heat dissipation without drastically deteriorating the κ value for graphene. Here also h-BN (which itself has a very high theoretically predicted thermal conductivity of 600 Wm$^{-1}$K$^{-1}$ [10] as opposed to experimentally reported value of ~140 Wm$^{-1}$K$^{-1}$ [17] in Si) has proved to be a better alternative as substrate to improve the κ and g values for graphene supported on h-BN compared to SiO$_2$/Si substrate.

Thermal studies reported so far show that the thermal conductivity of suspended [3,18] graphene gets suppressed from 5000 Wm$^{-1}$K$^{-1}$ [3] to 600 Wm$^{-1}$K$^{-1}$ for SiO$_2$/Si supported graphene [10,19]. Notably, the theoretically and experimentally predicted interface thermal conductance of graphene supported on SiO$_2$/Si varies in a wide range between 1-187 MWm$^{-2}$K$^{-1}$[20]. The interface thermal conductance of graphene supported on SiC and Au/SiO$_2$/Si has been reported as ~ 0.02 MWm$^{-2}$K$^{-1}$[21] and 28(+16/-9.2) MWm$^{-2}$K$^{-1}$ [22], respectively.

Another approach for dissipation of heat in graphene-based devices can be to replace the graphene channel by suitable van-der Waals heterostructures in the SiO$_2$/Si supported graphene-based nanoscale devices. Because of the superior performance of graphene devices when supported on h-BN substrate and as h-BN itself has a layered structure with similar lattice constant as graphene which may be exfoliated to extract 2D layers, h-BN and graphene promise to be a suitable pair for such heterostructure geometries [23-26] in devices. Zhang *et al.* theoretically showed that when graphene layer is supported on h-BN instead of SiO$_2$/Si, the thermal conductivity improves to 1347.3 ± 20.5 Wm$^{-1}$K$^{-1}$ [27] from 600 Wm$^{-1}$K$^{-1}$ [10]. Further, Liu *et al*. reported h-BN/graphene lateral heterostructure to show interfacial thermal conductance as high as 6.42×10$^9$ Wm$^{-2}$K$^{-1}$ [28]. A study of the thermal properties of graphene and its heterostructures with h-BN would enable us to better understand the performances of such graphene-based devices. To the best of our knowledge, there are very few experimental reports on the interfacial thermal conductance per unit area i.e., the g-values of van-der Waals heterostructures of graphene and h-BN supported on SiO$_2$/Si and, notably, those reports lack consistency [29,30] (See table S2 in Supporting Information). It may be noted that Yue *et al.* [31] have reviewed various experimental and theoretical Molecular Dynamics studies on thermal transport phenomena across the interfaces and thermal contacts in low dimensional devices. As discussed by Yue *et al.* [31], the understanding of the thermal properties involves a very precise measurement of the thermal contact resistance. As the traditional techniques of optical/electrical thermometry fail at nanoscale, the experimental works mostly depend on Raman thermometry (which is based on the scattering of the photon energy of the laser probe) for the accurate determination of the interfacial thermal resistance [31].

In this work, we have performed thermal studies on a series of van-der Waals (hetero)structures of graphene and h-BN to get a better understanding for potential applications in thermal management in nanoscale devices. We have prepared four different types of (hetero)structures: (a) h-BN (BN), (b) graphene (Gr), (c) h-BN on top of graphene (BG), and (d) graphene encapsulated by two h-BN layers from top and bottom (BGB), all supported on SiO$_2$/Si substrates. With the help of temperature- and power-dependent Raman spectroscopy, we have measured the thermal conductivity (κ) and interface thermal conductance per unit area (g) of the (hetero)structures by using an optothermal method – a Raman spectroscopy-based technique [3]. In comparison to graphene, we find an 80% increase in κ and twenty times increase in g in the BG-heterostructure while the BGB-heterostructure enhances the g by almost two orders of magnitude.

**Experimental details:**

**Preparation of (hetero)structures:**

Graphene monolayer and few-layers hexagonal boron nitride (h-BN) flakes were mechanically exfoliated on Si substrate coated with ~ 300 nm thick $SiO_2$ film using scotch tape technique from the bulk HOPG graphite and bulk h-BN crystals, respectively. The heterostructures (BG and BGB) were prepared by stacking the exfoliated layers using a micromanipulator by "hot pickup" technique [32,33]. To prepare the BG heterostructure, a h-BN flake was picked up at 45$^o$C using a Polypropylene carbonate (PPC) coated polydimethylsiloxane (PDMS) block mounted on a glass slide attached to the tip of a micromanipulator. The h-BN flake was then transferred on top of the previously exfoliated graphene monolayer at 75$^o$C on $SiO_2$/Si (p++) substrate. The BG heterostructure was then cleaned in chloroform ($CHCl_3$) followed by acetone and isopropyl alcohol (IPA). Similarly, to prepare the BGB heterostructure, a h-BN flake was picked up as per the process discussed above. This h-BN flake was then aligned on top of an exfoliated monolayer graphene. Now, the graphene monolayer was picked up using the h-BN at 50$^o$C. The h-BN/graphene stack was then aligned and dropped down on another exfoliated h-BN flake at 75$^o$C. Finally, the BGB stack was cleaned in chloroform ($CHCl_3$) followed by acetone and isopropyl alcohol (IPA). The steps followed for the fabrication of the BG and BGB heterostructures are shown in Figure 1(a).

**Measurements:**

The thickness of the various samples was determined using Raman and AFM measurement (see supplementary note S1 and supplementary Figures S1 and S2 in Supporting Information). The graphene layers are found to be monolayers while the h-BN was about ~ 10 layers (~3.4 nm) thick. Raman measurements were performed in a back-scattering geometry by using LABRAM HR Evolution Raman spectrometer (1800 rulings/mm), fitted with a microscope with 50X (NA=0.5) and 20X (NA=0.4) objectives, coupled to an air-cooled charge coupled device detector. All the measurements were carried out with 532 nm laser excitation. The pixel resolution was ~ 0.37 $cm^{-1}$. In order to measure the laser spot size for different objectives, we have performed a Raman linear mapping across a sharp-edged Au strip deposited on $SiO_2$/Si substrate, as shown in Figure 1(b, c). The spot radii were determined to be 0.64 μm (for 50X objective lens) and 0.94 μm (for 20X objective lens), see supplementary note S2 in Supporting Information for details. The temperature-dependent Raman studies were performed using fixed laser power (~1 mW) inside a LINKAM heating stage (Model HFS600E-PB4) and varying the temperatures from 300 to 450 K in steps of 10 K. All the power dependent measurements were performed at room temperature varying the laser power using appropriate neutral density filters.

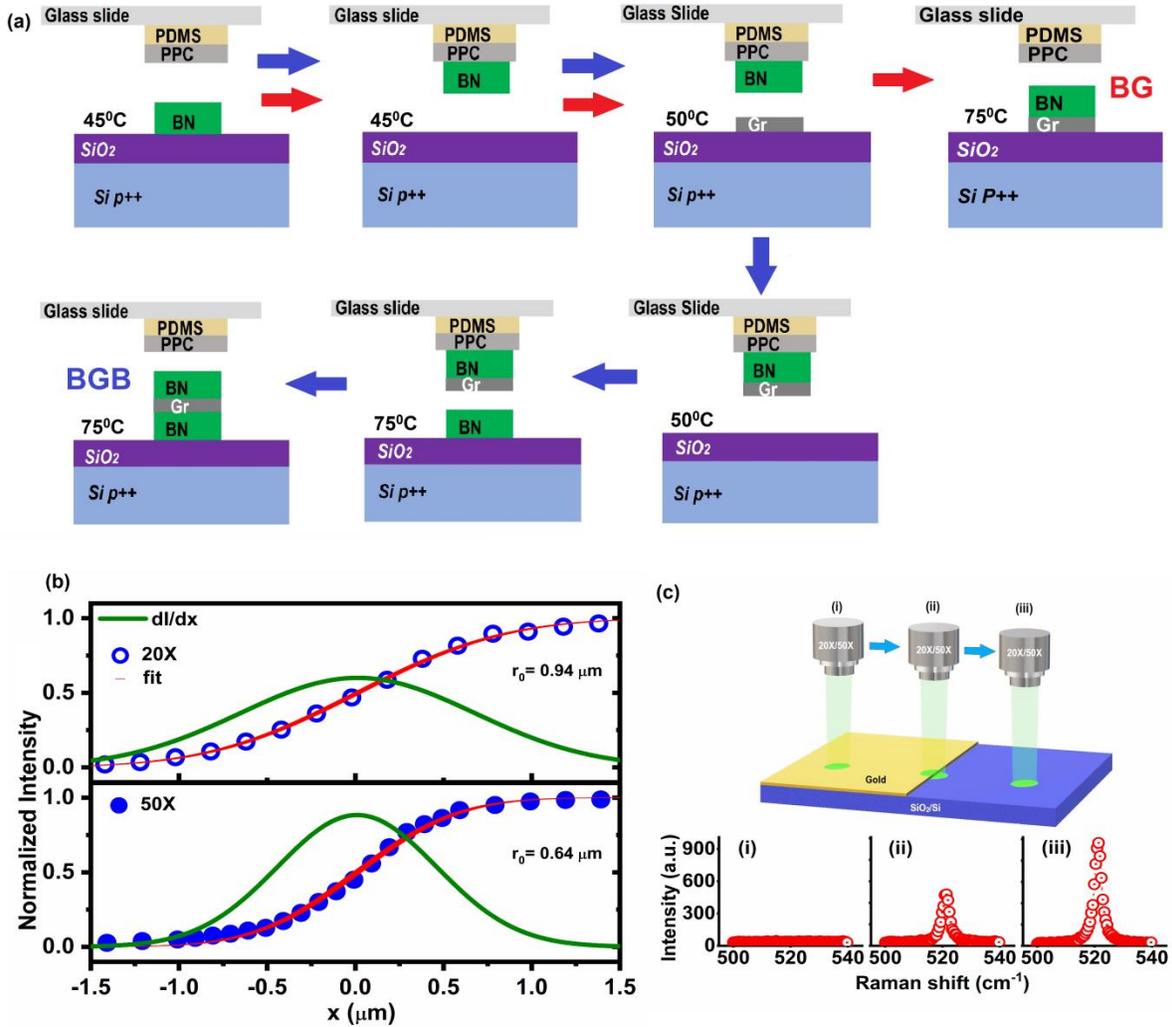

*Figure 1*. *(a) Schematic of the fabrication process of the BG and BGB heterostructures. Direction of red (blue) arrows shows the sequence of processes in the fabrication of BG (BGB) heterostructure. (b) Normalized intensity profiles of the silicon (Si) Raman mode at 520 $cm^{-1}$ along the linear map taken across a sharp gold strip deposited on $SiO_2$/Si substrate. The solid and open circles represent the data for 50X and 20X objective lenses, respectively. Red curves show the fitting of the normalized intensity (I) profiles and green curves show the dI/dx, (c) A schematic representing the technique employed to measure the laser spot size. The plots (i), (ii), and (iii) show the Si Raman mode at different positions of the laser spot.*

## Results and Discussion

Figure 2(a, b) shows the schematic diagrams and corresponding optical images of the four different (hetero)structures – BN, Gr, BG, and BGB. The room temperature Raman spectra for the different

(hetero)structures are shown in Figure 2(c). The absence of D (defect) band in the Raman spectra signifies that the graphene layers used were pristine and defect free. The line shape of the 2D band (which can be fitted with a single Lorentzian function for monolayer graphene) [34] and the relative intensity of 2D band with respect to the G band confirm that all the graphene flakes used in the (hetero)structures were monolayers, see Figure S1 in Supporting Information. AFM measurements reveal the thickness of the h-BN flakes used to be ~10 layers (See Supplementary Note 1(B) of Supporting Information). From the temperature- and power-dependent Raman spectroscopic measurements, it was observed that the G band was less sensitive to temperature and laser power than the 2D band. Therefore, all the analyses have been performed using 2D band as was also previously done by Chen *et al*. [35] and Jae-Ung Lee *et al*. [36].

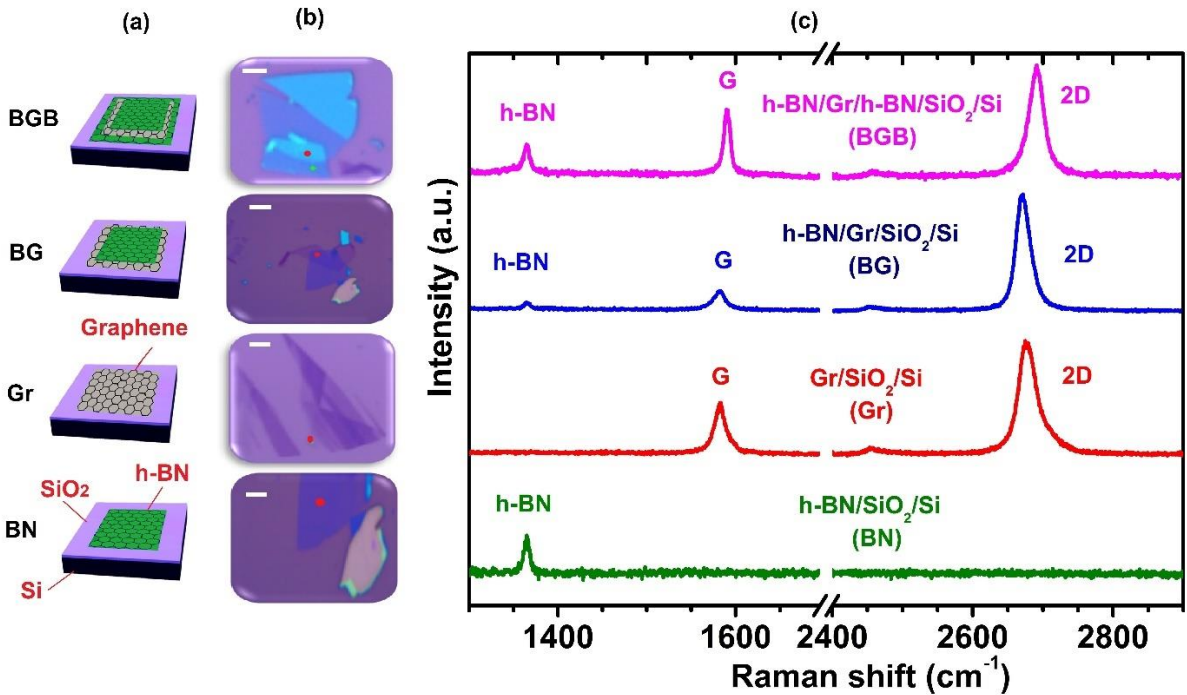

*Figure 2. (a) Schematic diagram showing the four different samples: BN, Gr, BG, BGB with labeled components; (b) Optical images of the four different samples taken with 50X objective. The red dots represent the investigated area of the samples. The white bars are the scales corresponding to 10 μm; (c) Corresponding Raman spectra of the four different samples.*

Figure 3(a) shows the color maps representing the normalized intensity profile of the 2D mode (for Gr, BG, and BGB) and h-BN mode (for BN) across the complete temperature range where we can clearly observe a redshift of both the modes with an increase in the temperature, see Figure S3 (a) in Supporting Information. We have observed a similar redshift of the modes with an increase in laser power, as shown in Figure 3(b) and Figure S3 (b) in Supporting Information. This is because of the local heating of the sample by the laser. The redshift of the modes can be explained by thermally induced softening of bonds as was previously reported by Calizo *et al.* [37]. The temperature- and power-dependences of the 2D and h-BN modes of the different (hetero)structures are shown in Figure 4(a-h) and the corresponding first order temperature- and power- coefficients, $\chi_T = \left(\frac{d\omega}{dT}\right)_P$ and $\chi_P = \left(\frac{d\omega}{dP}\right)_T$, respectively, are enlisted in Table 1.

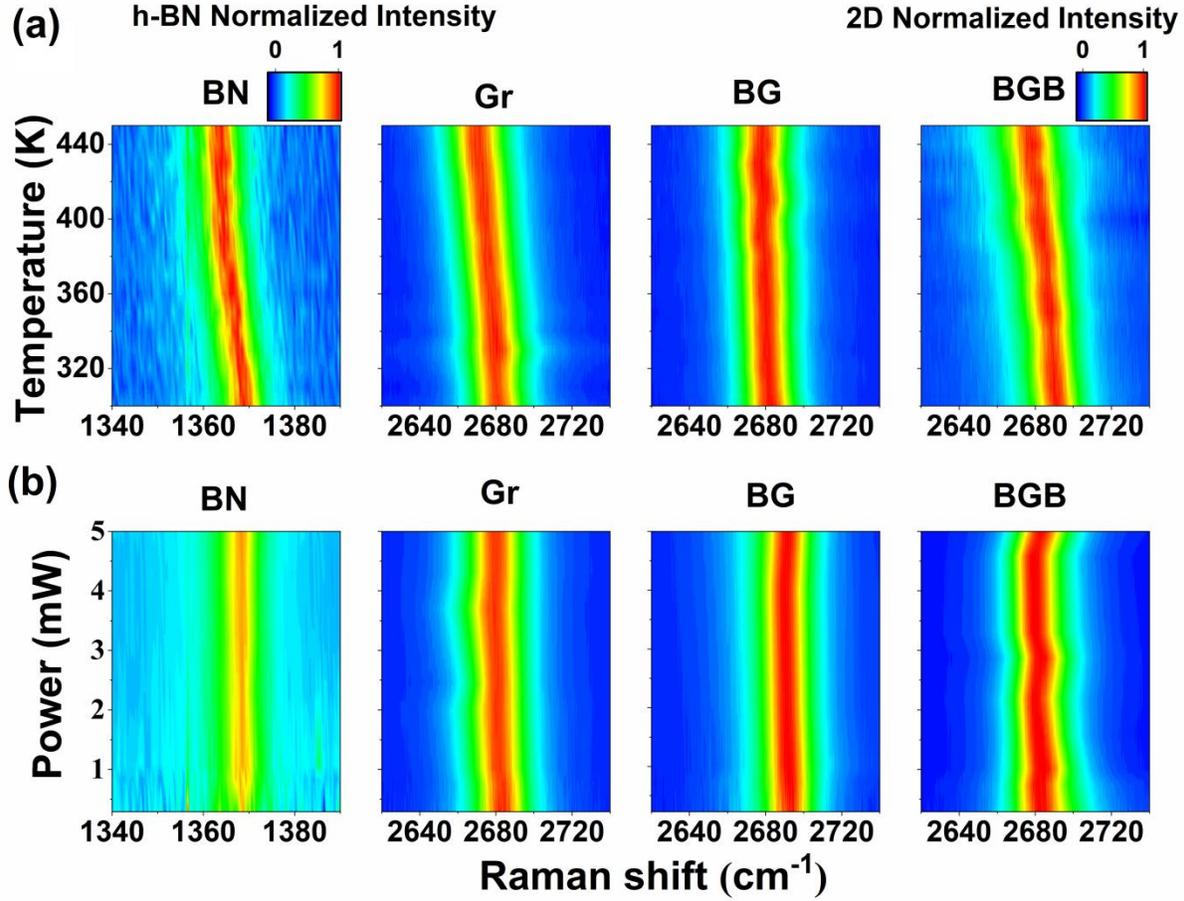

*Figure 3. Color maps representing the normalized intensity profile as a function of (a) temperature vs Raman shift and (b) power vs Raman shift for the 2D mode (for Gr, BG, BGB samples) and h-BN mode (for BN sample)*

In order to determine the κ and g values of the (hetero)structures, we have considered the following heat flow equation (Eq 1) assuming no heat loss to the surrounding atmosphere [22] (see supplementary note S4 in Supporting Information):

$$\frac{1}{r}\frac{d}{dr}\left(r\frac{dT}{dr}\right) - \frac{g}{\kappa l}(T - T_a) + \frac{q}{\kappa} = 0 \qquad (1)$$

Here T, $T_a$, g, κ, $l$, and q represent the temperature at position r, ambient temperature, interface conductance, thermal conductivity, thickness of the (hetero)structure, and volumetric heat coefficient, respectively. The above equation can be used to estimate the interfacial thermal resistance $R_m$ (see supplementary note S4 in Supporting Information) which is related to experimentally obtained $χ_P$ and $χ_T$ as:

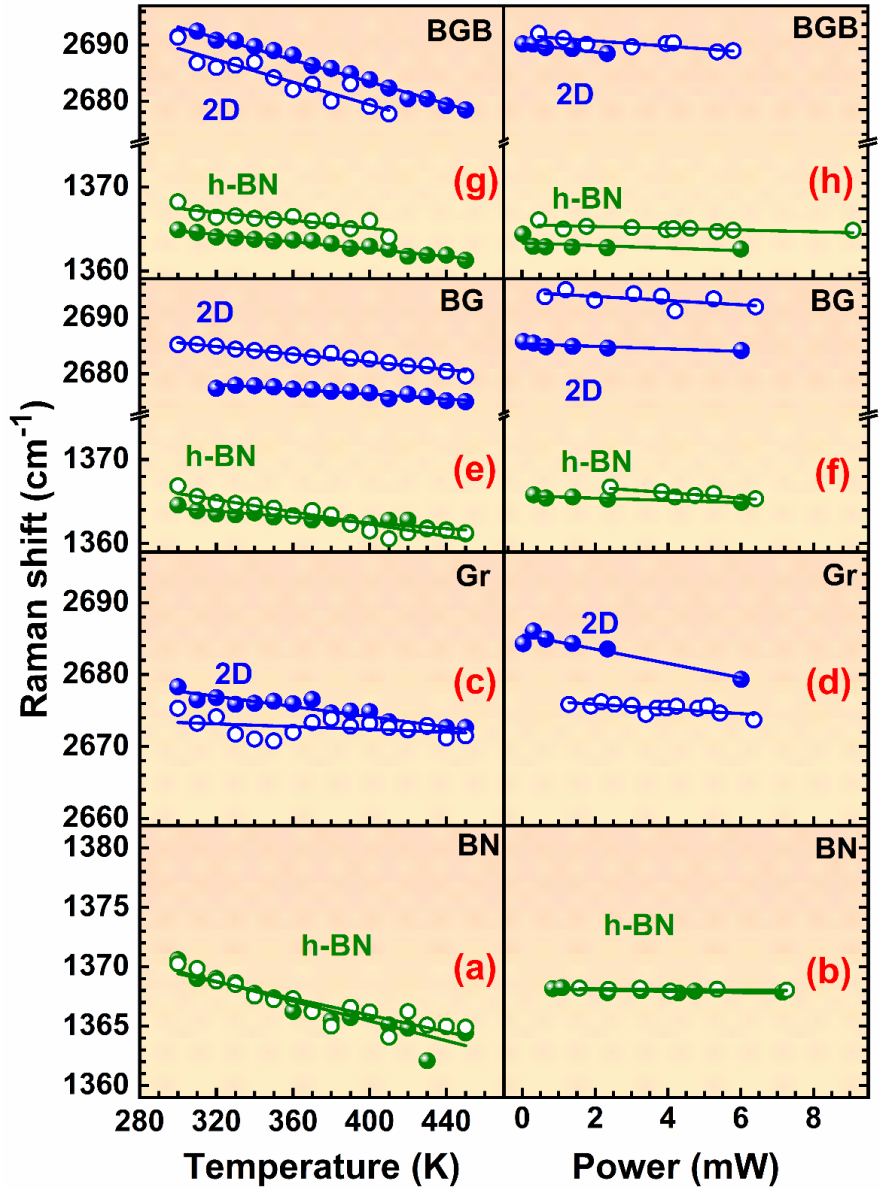

*Figure 4.* *Temperature- and power- dependence of 2D and h-BN Raman modes in: (a,b) BN sample, (c,d) Gr sample, (e,f) BG sample, (g,h) BGB sample respectively. The slopes of the plots give the first order temperature and power coefficients which are enlisted in Table 1. Filled circles represent the data recorded by 50X objective while the open circles correspond for 20X objective.*

**Table 1:** The first order temperature and power coefficients ($\chi_T$ and $\chi_P$) of the investigated (hetero)structures.

| Sample | Mode | Objective | $\chi_T$ | $\chi_P$ |
|---|---|---|---|---|
| Gr | 2D | 50X | -0.035 cm$^{-1}$/K | -0.993 cm$^{-1}$/mW |
| Gr | 2D | 20X | -0.009 cm$^{-1}$/K | -0.308 cm$^{-1}$/mW |
| BN | h-BN | 50X | -0.041 cm$^{-1}$/K | -0.052 cm$^{-1}$/mW |
| BN | h-BN | 20X | -0.034 cm$^{-1}$/K | -0.025 cm$^{-1}$/mW |
| BG | 2D | 50X | -0.022 cm$^{-1}$/K | -0.226 cm$^{-1}$/mW |
| BG | 2D | 20X | -0.034 cm$^{-1}$/K | -0.373 cm$^{-1}$/mW |
| BG | h-BN | 50X | -0.017 cm$^{-1}$/K | -0.134 cm$^{-1}$/mW |
| BG | h-BN | 20X | -0.036 cm$^{-1}$/K | -0.316 cm$^{-1}$/mW |
| BGB | 2D | 50X | -0.098 cm$^{-1}$/K | -0.671 cm$^{-1}$/mW |
| BGB | 2D | 20X | -0.100 cm$^{-1}$/K | -0.437 cm$^{-1}$/mW |
| BGB | h-BN | 50X | -0.022 cm$^{-1}$/K | -0.203 cm$^{-1}$/mW |
| BGB | h-BN | 20X | -0.024 cm$^{-1}$/K | -0.137 cm$^{-1}$/mW |

$$R_m = \frac{\partial \theta_m}{\partial Q} = \frac{\partial \omega}{\partial Q}\frac{\partial \theta_m}{\partial \omega} = \chi_P(\chi_T)^{-1} \qquad (2)$$

Using the $R_m$ value obtained from equation (2), we have measured the g and κ of the different (hetero)structures.

Figure 5 shows the measured g and κ values for the (hetero)structures. The g and κ values of h-BN supported on SiO$_2$/Si (BN) are 25.6±0.4 MWm$^{-2}$K$^{-1}$ and 280.0±58.0 Wm$^{-1}$K$^{-1}$, respectively. We have observed a lower value of κ for h-BN as compared to the theoretical report on freestanding h-BN [14] due to the substrate effect of SiO$_2$/Si. However, the κ for h-BN is still appreciably high and, hence, h-BN can be used as a channel for heat dissipation. We report the g for graphene supported on SiO$_2$/Si (Gr) as 1.15±0.40 MWm$^{-2}$K$^{-1}$ which is about an order higher than the g reported for graphene supported on SiC [21]. Our measured value of κ for Gr (600.0±93.0 Wm$^{-1}$K$^{-1}$) is consistent with previously reported values.

Importantly, the effective g and κ values for the heterostructure BG are 26.2±0.8 MWm$^{-2}$K$^{-1}$ and 1072±88 Wm$^{-1}$K$^{-1}$, respectively, while for the BGB heterostructure the values are 105±1 MWm$^{-2}$K$^{-1}$ and 850.0±81.0 Wm$^{-1}$K$^{-1}$, respectively. The uncertainties in the measured κ and g values for the different (hetero)structures have been calculated on the basis of the values obtained from multiple sets of temperature- and power-dependent Raman measurements performed on the various (hetero)structures.

It has been theoretically predicted that in order to have a finite κ, the hexagonal 2D crystals must follow the relation [8, 38]:

$$\omega_{qs}^2 n_{qs}^0 (n_{qs}^0 + 1) v_{qs}^2 \tau_{qs}^{ph} \propto q^n \quad \text{with } n > -2 \quad (3)$$

Due to the high phonon density of states and selection rule by Lindsay *et al.* [39], the ZA phonons are predicted to be the principal contributors to the thermal conductivity of graphene. The ZA phonons show a quadratic dispersion, and hence has a phonon lifetime that varies as $q^n$ with $n > -4$. The ZA phonons are very sensitive to strain and even a slight strain can cause a linearization of the ZA phonon dispersion resulting in a reduction in κ. For graphene supported on SiO$_2$/Si, a complete renormalization of the ZA phonon dispersion takes place leading to an upshift of the ZA phonon mode at the *Γ* point of the Brillouin zone [40-42]. This results in the strong suppression of the κ value for the SiO$_2$/Si supported graphene (Gr).

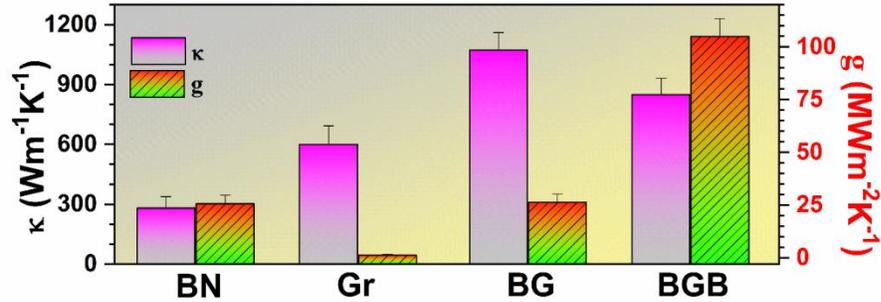

***Figure 5.*** *Bar diagram comparing the κ (left axis) and g (right axis) values for the different samples.*

It has been reported that h-BN substrate has a considerably negligible influence on the dispersion of graphene but it strongly suppresses the flexural modes [43]. Since the lifetime of flexural phonons is strongly suppressed for graphene supported on h-BN, the lifetime of in-plane phonons increases. It has also been reported that the contribution of the flexural acoustic phonons (ZA) to the thermal conductivity for heterostructures of h-BN and graphene is lesser than that of the in-plane phonon modes (LA, TA) [44]. Further, Zou *et al.* [43] reported that there is a significant contribution in thermal conductivity of graphene/h-BN heterostructures from the optical phonons as well. The relatively larger phonon mean free path for the h-BN supported graphene, as compared to the SiO$_2$/Si supported case, results in an effective increase of its thermal conductivity. This is consistent with our results as the heterostructures of graphene and h-BN show an impressive increase in the thermal conductivities by ~ 79% (for BG) and ~ 42% (for BGB) with respect to Gr. The slight decrease in the κ value for BGB as compared to BG can be attributed to greater suppression of the flexural modes [45].

The cross-plane interfacial conductance depends on two factors: (a) the effective area of contact at the interface, and (b) the in-plane thermal conductivity of the substrate [45, 46]. The effective contact area between the graphene channel and the substrate increases with the increasing flatness of the surface that graphene is exposed to. The SiO$_2$/Si substrate normally has a rough surface [47-49] (that plays an important role on the properties of graphene) and, hence, the actual contact area between the substrate and graphene layers is much lesser than the case when a layer of h-BN is introduced as a buffer layer between the graphene and SiO$_2$/Si substrate (see Figure S4 in Supporting Information). An increased contact area ensures increased number of channels for heat dissipation and hence an increase in the interfacial thermal conductance per unit area (g) [50-52]. This can explain the increasing trend of g with the increase in the number of h-BN and graphene interfaces [53], as seen in Figure 5. The h-BN lattice, due to its excellent lattice matching with graphene provides a flat surface which increases the effective contact area at the interfaces [54]. The g value we report for the BG sample is ~ 3.5 times the value reported by Chen *et al.* [29]. Finally, the BGB heterostructure shows the value of g which is orders of magnitude higher than the monolayer graphene (Gr) and the heterostructure BG. This is because here the central graphene channel is sandwiched from both (top and bottom) sides by the flat h-BN layers and hence the effective contact area at the interfaces is increased to a greater extent. The other important factors are: (i) the h-BN on SiO$_2$/Si (BN) itself shows a greater g value compared to Gr, so we may conclude that h-BN layer dissipates the heat more effectively through the SiO$_2$/Si substrate as compared to graphene, (ii) the substantially high κ value of h-BN (though lesser than graphene) ensures good in-plane thermal transport. It must be noted here that the thermal properties of the investigated heterostructures strongly depend on the local in-plane and out-of-plane forces acting on the carbon atoms in the graphene sheets [45]. The interlayer coupling with h-BN has been reported to strongly influence the local forces acting on the graphene sheet. The excellent lattice matching with h-BN and the strong interlayer adhesion forces induced by sublattice charge polarization along the graphene lattice, result in a reduction of graphene sheet corrugation and the local in-plane strain forces. This in turn enhances the in-plane phonon lifetimes, which are the principal contributors to the thermal transport properties in h-BN/graphene heterostructures. The addition of more h-BN layers is expected to reduce the in-plane strain forces, thereby enhancing the in-plane phonon lifetimes further. The special case of BGB heterostructure, where the graphene layer is sandwiched by h-

BN from top and bottom, has been theoretically predicted to show the strongest cross-planar adhesion, which in turn results in excellent surface flatness and minimal local strain forces on the graphene sheet [45]. These factors together contribute to the excellent heat dissipation properties of the BGB heterostructure. The h-BN layer thickness dependence of the obtained κ and g values for the heterostructures are further discussed in supplementary note S6 of the Supporting Information.

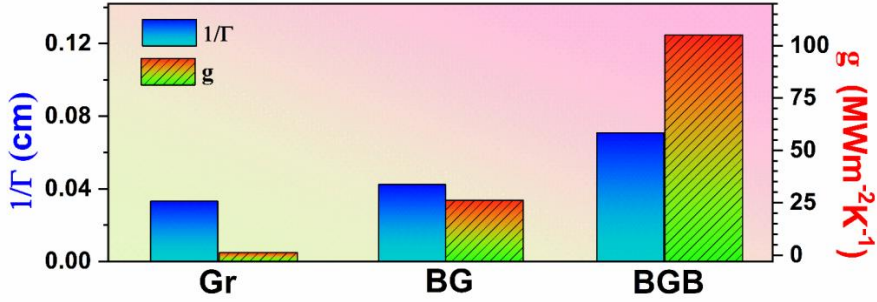

*Figure 6. Bar diagram showing the inverse linewidth of the 2D mode of graphene (left axis) and interface conductance per unit area g of the samples (right axis).*

The observed dependence of the κ and g in the (hetero)structure samples may also be understood by investigating their phonon lifetime. The linewidth ($\Gamma$) of a phonon mode in a Raman spectrum is inversely related to the lifetime (τ) of the phonon (i.e., $\Gamma \propto \frac{1}{\tau}$) [55] (see supplementary Figure S5 in Supporting Information). Though the major contribution to thermal conductivity in graphene/h-BN heterostructures comes from the in-plane and out-of-plane acoustic modes, recent reports [43] show that the optical modes also contribute significantly. We, therefore, expect to see similar trends in the acoustic and the optical mode lifetimes. Figure 6 shows the inverse linewidth for the 2D modes of graphene in the different (hetero)structures which are compared with the corresponding g-values. It can be seen that the inverse linewidth of the 2D band (hence the lifetime of the associated phonon) also shows an increasing trend with the increase in the number of h-BN-graphene interfaces, similar to the measured g-values. This can be attributed to the excellent lattice matching of h-BN with graphene which ensures a flatter surface for the graphene layer thus increasing the lifetime of the in-plane phonons [45]. As previously discussed, the flatness of the surface to which graphene is exposed increases the interfacial conductance. Further, as the phonon lifetime is directly proportional to the phonon mean free path, by using the equation (4) [45], we would expect an improved thermal conductance (κ) for the h-BN and graphene heterostructures, as has been seen in our data,

$$\kappa = \frac{1}{3} C_v \lambda v \qquad (4)$$

where $C_v$, λ, and $v$ are the specific heat capacity at constant volume, phonon mean free path, and phonon group velocity respectively.

*Conclusion:*

We have performed a systematic study of thermal properties of graphene, h-BN, and two of their heterostructures (BG: h-BN/graphene, and BGB: h-BN/graphene/h-BN configurations), all supported on SiO$_2$/Si substrate, using Raman spectroscopy. As Raman spectroscopy is a non-destructive and sensitive characterizing tool for 2D materials, it makes the study of the thermal properties of 2D (hetero)structures very convenient. The thermal conductivity (κ) and the interfacial thermal conductance per unit area (g) were measured for the (hetero)structures of graphene and h-BN. The κ of graphene supported on SiO$_2$/Si is found to be comparable to the reported values while that for h-BN on SiO$_2$/Si substrate was observed to be appreciably high. We have also measured the g for the different (hetero)structures and observed that for BN it was greater than Gr, which indicates that the cross-plane heat dissipation through SiO$_2$/Si substrate is greater for h-BN as compared to graphene. The g for the heterostructures is higher and it was found to be maximum for the BGB heterostructure because of the higher flatness of the h-BN layers that the graphene channel was exposed to. The flat surface of the h-BN ensures more contact area and more channels for heat dissipation and, hence, a higher g (see Figure S4 in Supporting Information). Based on our results, we conclude that because of the superior thermal performance (a considerably less suppression of κ along with very high g) of the BGB heterostructure, h-BN encapsulated graphene can offer a better alternative for heat dissipation in nanoscale devices.

*Acknowledgements:*

We sincerely acknowledge the funding from DST/SERB (Grant No. ECR/2016/001376 and Grant No.CRG/2019/002668), Nanomission (Grant No.SR/NM/NS-84/2016(C)), and DST-FIST (Grant No. SR/FST/PSI-195/2014(C)). The authors acknowledge the Central Instruments Facility, IISER Bhopal, for AFM facility.

**Supporting Information:** Determination of flake thickness using AFM and Raman data, determination of laser spot size, temperature- and laser power-dependent Raman spectra, discussion on the Optothermal method with detailed calculations for different (hetero)structures, qualitative discussions regarding the dependence of thermal properties on roughness, flake thickness and phonon lifetime.

# SUPPORTING INFORMATION

# Hexagonal Boron Nitride-Graphene Heterostructures with Enhanced Interfacial Thermal Conductance for Thermal Management Applications

Saheb Karak[1], Suvodeep Paul[1], Devesh Negi[1], Bommareddy Poojitha[1],
Saurabh Kumar Srivastav[2], Anindya Das[2], and Surajit Saha[1*]

[1]Department of Physics, Indian Institute of Science Education and Research, Bhopal, 462066, India

[2]Department of Physics, Indian Institute of Science, Bangalore, 560012, India

*Correspondence: surajit@iiserb.ac.in

**Table of Contents:**

S1. Detection of number of graphene and h-BN layers in the samples under study

S2. Determination of laser spot size

S3. Temperature- and Laser-power- dependence of the Raman spectra

S4: The Optothermal method with calculations and theoretical background

S5. Dependence of surface contact area and surface roughness on introducing h-BN

S6. Effect of flake thickness on $\kappa$ and $g$

S7. Phonon lifetime vs linewidth of Raman modes

Table S1.

Table S2.

**S1. Detection of number of graphene and h-BN layers in the samples under study**

A.      The samples used for the measurements here are comprised of monolayer graphene. Sample sizes vary from ~ 10-30 μm which are far larger than the laser spot size (~ 1 μm). Chain

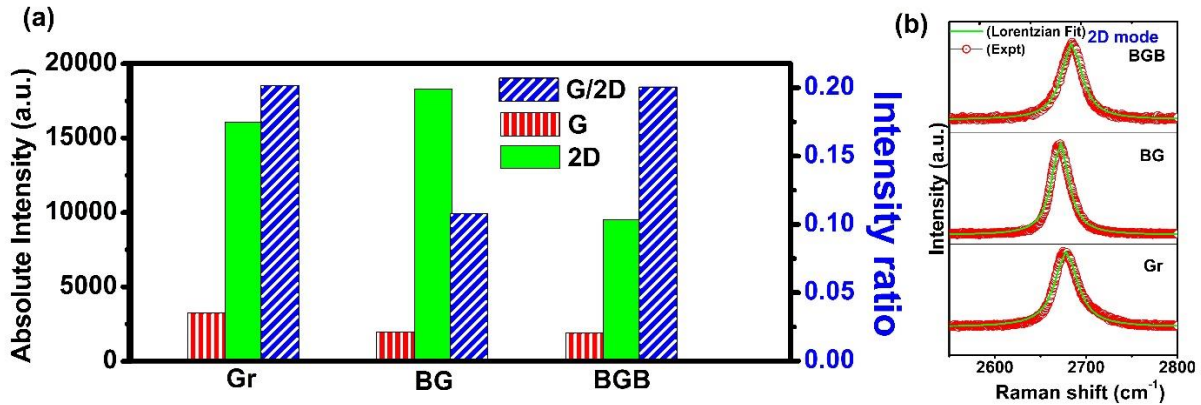

*et al*. reported that the graphene size effect on thermal conductivity is suppressed in supported graphene [1]. The layer number was verified from Raman spectra of the samples at room temperature as shown in original text Figure 2(c). It was observed that the absolute intensities of the G mode were lower than those of the 2D mode in the Raman spectra for all the samples with $I_G/I_{2D} < 0.5$. This has been reported to be a signature of monolayer graphene [2-4]. Figure S1(a) shows the absolute intensities of the G and 2D modes and their relative ratios (which varies in the range of 0.1 to 0.2) for the different samples thus confirming the graphene to be of monolayer thickness. Further, the 2D bands in the Raman Spectra of monolayer graphene has been reported to show single Lorentzian line shape, while multilayer graphene shows a line shape which may be fitted with multiple Lorentzian functions [3]. Figure S1(b) clearly shows that the 2D bands of the graphene flakes in Gr, BG, and BGB (hetero)structures are fitted with single Lorentzian functions, again confirming the graphene flakes to be of monolayer thickness.

*Figure S1*: *(a) Bar diagram showing the absolute intensity of G and 2D Raman modes (left axis) and ratio of intensities of G and 2D mode (right axis). (b) 2D band of graphene in Gr, BG, BGB (hetero)structures.*

As has been previously reported, the G and 2D band intensity ratio also depends on charge doping [5] and hence the differential value observed in different samples in use may be due to different extents of unintentional charge doping that may have occurred in the various samples. However, as the intensity ratio values range between 0.10 - 0.20 for our samples, we may safely speculate that the small variations in the unintentional charge doping in the different samples will not have any consequence on the properties investigated in this work. The following table elaborates the observations and confirms that the graphene in our samples is monolayer.

**Table S1:** Area under the curve values for G and 2D modes obtained by fitting the room temperature Raman spectra of the related (hetero)structures using Lorentzian functions.

| (Hetero)structures | $I_G$ [G mode Area under the curve (unit$^2$)] | $I_{2D}$ [2D mode Area under the curve (unit$^2$)] | $I_G/I_{2D}$ |
|---|---|---|---|
| Gr | 3243 | 16073 | 0.2 |
| BG | 1977 | 18304 | 0.1 |
| BGB | 1909 | 9519 | 0.2 |

B.  The thicknesses of the h-BN flakes in the heterostructures were determined from AFM measurements in contact mode (Agilent 5500). It is well known that the thickness of monolayer h-BN is ~ 0.333 nm [6]. The Figure S2 shows the AFM topography image along with the height profile revealing the thickness of the h-BN flakes to be ~ 10 layers [6]. It is worth mentioning that supported 2D layers on a substrate improves their thermal dissipation properties with an increase in the number of layers. Therefore, fabricating a heterostructure (of graphene and h-BN) with multilayer h-BN is expected to improve the thermal performance with respect to the heterostructure with a single h-BN layer.

*Figure S2: Left: AFM image showing the topography of h-BN layer. Right: Height profile along the white solid line in the left image.*

## S2. Determination of laser spot size

In order to determine the laser spot radius, a technique similar to the knife-edge method was employed [7]. A sharp-edged thin strip of gold was deposited on the $SiO_2$/Si substrate and then a linear Raman mapping was done across the sharp edge of the gold strip in the range of 450 cm$^{-1}$ to 550 cm$^{-1}$. The Figure 1(b) shows the linear mapping measurement done. The Raman spectra obtained at the different positions marked (i), (ii), and (iii) are shown and we could clearly observe the disappearance of the Si peak at 520 cm$^{-1}$ as the laser spot was moved across the sharp gold strip. Figure 1(c) shows the corresponding intensity profile of the Si peak along the linear map. Figure 1(b), the intensity profile was fitted with the function:

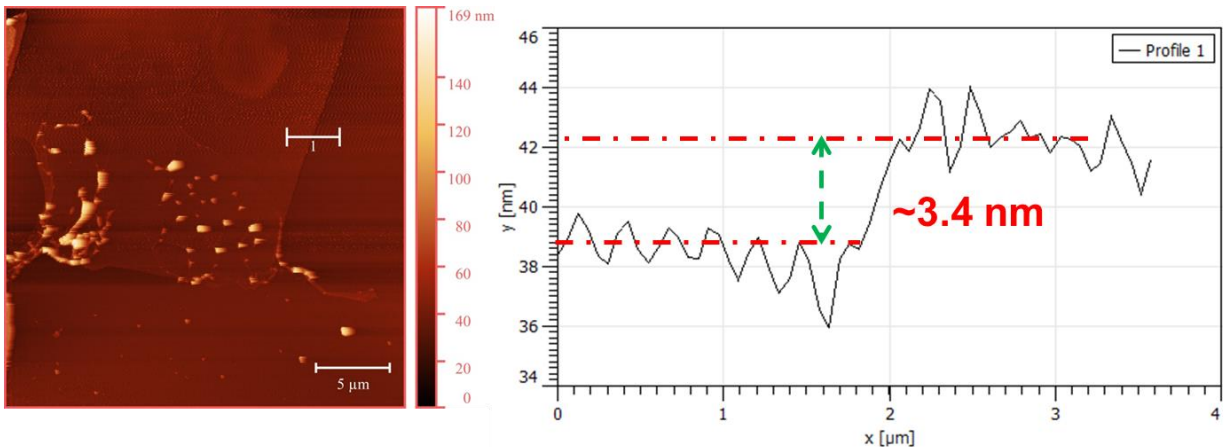

$$I(x) = \frac{I_0}{2}\left(1 + \mathrm{erf}\left(\frac{x-x_0}{w_0}\right)\right)$$

The $\frac{dI}{dx}$ for the fitted intensity profile showed a clear Gaussian behaviour and was fitted by the function of type $\exp\left(\frac{-x^2}{r_0^2}\right)$ and hence the spot radius $r_o$ for 20X (0.94μm) and 50X (0.64μm) objective lenses were determined.

### S3. Temperature- and Laser-power-dependence of the Raman spectra

Temperature and power dependent Raman studies were done for all the samples. Figure S3 (a,b) shows the stacks of temperature and power dependent spectra for the different samples. It was observed that all modes show a red-shift with increase in temperature as well as power. This redshift can be attributed to phonon-phonon interaction mediated anharmonic contributions to the interatomic Coulombic potential energy at higher temperatures [8]. The power dependent spectra also show similar red-shifts because the laser-power induces local heating of the sample thereby increasing the temperature near the exposed part of the sample. We have performed laser power dependent experiments in a range from 0.03 mW – 7.15 mW, which results in a huge enhancement in the intensity of the Raman spectra for the various (hetero)structures with increasing laser power thus lowering the noise level in the spectra recorded at higher powers. On the other hand, the temperature-dependent measurements are performed at a fixed laser power (~1 mW). Hence, the signal-to-noise ratio in the temperature-dependent spectra remains almost comparable except at very high temperature say at 450 K. At around 450 K, the adhesive residues that are present on the substrate due to the micromechanical exfoliation process start to evaporate and that often cause a degradation of the counts (or intensity) at high temperatures.

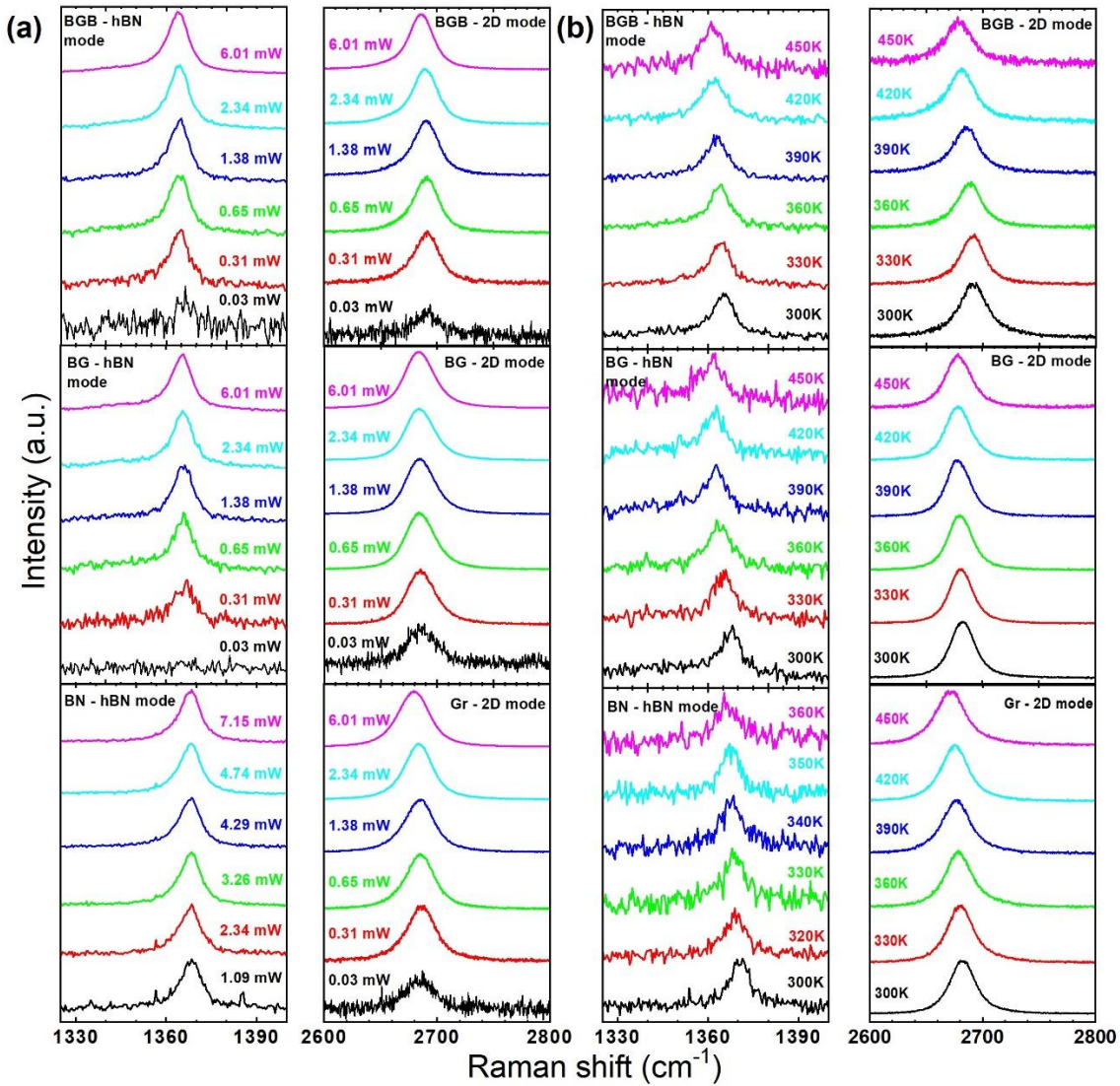

*Figure S3: (**a**) Comparison of temperature-dependent and (**b**) power-dependent Raman spectra showing the h-BN mode (left) and 2D mode (right) of the different samples. Both the modes show redshift with increase in temperature (and power).*

**S4: The Optothermal method with calculations and theoretical background**

The optothermal method is an easy and non-destructive approach of measuring the thermal conductivity ($\kappa$) and interfacial thermal conductance (g) of a material supported on a given substrate, wherein temperature- dependent and power-dependent Raman response is measured. This technique has often been used for the measurement of κ and g of various 2D layered materials [9]. Therefore, we have used this technique for the measurement of κ and g of our samples.

The thermal conductivity (κ), and interface thermal conductance per unit area (g) for the different samples on SiO$_2$/Si substrate were estimated by using the Raman spectroscopy- based measurement using the approach developed by Cai *et al*. [9]

A 2D-layer when supported on a substrate can dissipate the heat through the in-plane and out-of-plane heat conduction channels, (neglecting the contribution from convection and radiation losses). The temperature distribution in the sample can be obtained from heat diffusion equation in cylindrical coordinates as follows [9-11]

$$\frac{1}{r}\frac{d}{dr}\left(r\frac{dT}{dr}\right) - \frac{g}{\kappa l}(T - T_a) + \frac{q}{\kappa} = 0 \qquad (1)$$

Here T, T$_a$, g, κ, $l$, and q represent the temperature at position r, ambient temperature, interface conductance, thermal conductivity, thickness of the sample, and volumetric heat coefficient respectively. The volumetric optical heating q can be represented as:

$$q = \frac{q_0}{l}\exp\left(-\frac{r^2}{r_0^2}\right) \qquad (2)$$

Here $q_0$ and $r_0$ represent the peak absorbed laser power per unit area at the centre of the beam spot and the radius of the laser beam spot respectively. The total absorbed laser power $Q$ is then

$$Q = \int_0^\infty q_0 \exp\left(-\frac{r^2}{r_0^2}\right) 2\pi r\, dr = q_0 \pi r_0^2 \qquad (3)$$

Substituting $\theta \equiv (T - T_a)$ and $z = \left(\frac{g}{\kappa l}\right)^{\frac{1}{2}} r$ into equation (1), we get a nonhomogeneous Bessel's equation:

$$\frac{d^2\theta}{dz^2} + \frac{1}{z}\frac{\partial\theta}{\partial z} - \theta = -\frac{q_0}{g}\exp\left(-\frac{z^2}{z_0^2}\right) \qquad (4)$$

The solution to equation (4) is given as

$$\theta(z) = C_1 I_0(z) + C_2 K_0(z) + \theta_p(z) \qquad (5)$$

where $I_0(z)$ and $K_0(z)$ are the zero-order modified Bessel functions of first and second kind, respectively.

$$\theta_p(z) = I_0(z)\int K_0(z)\frac{\pi q_0}{2g}\exp\left(-\frac{z^2}{z_0^2}\right)dz - K_0(z)\int I_0(z)\frac{\pi q_0}{2g}\exp\left(-\frac{z^2}{z_0^2}\right)dz \qquad (6)$$

The boundary conditions $\left(\frac{d\theta}{dz}\right)_{z=0} = 0$ and $\theta(z \to \infty) = 0$ yield $C_2 = 0$ and $C_1 = -\lim_{z \to \infty}\left(\frac{\theta_p(z)}{I_0(z)}\right)$, which approach a constant value for large z.

The temperature rise in the sample measured by the Raman laser beam is given by:

$$\theta_m = \frac{\int_0^\infty \theta(r)\exp\left(-\frac{r^2}{r_0^2}\right)rdr}{\int_0^\infty \exp\left(-\frac{r^2}{r_0^2}\right)rdr} \quad (7)$$

We define the measure thermal resistance as $R_m \equiv \frac{\theta_m}{Q}$. On the basis of equations (3) and (7)

$$R_m = \frac{\int_0^\infty \left(-I_0(z)\lim_{z\to\infty}\frac{\theta_p(z)}{I_0(z)} + \theta_p(z)\right)\exp\left(-\frac{r^2}{r_0^2}\right)rdr}{\int_0^\infty \exp\left(-\frac{r^2}{r_0^2}\right)rdr \int_0^\infty q_0\exp\left(-\frac{r^2}{r_0^2}\right)2\pi rdr} \quad (8)$$

*For measurements with 50X objective:*

$$R_{m(50X)} = \frac{\int_0^\infty \left(-I_0(z)\lim_{z\to\infty}\frac{\theta_p(z)}{I_0(z)} + \theta_p(z)\right)\exp\left(-\frac{r^2}{r_1^2}\right)rdr}{\int_0^\infty \exp\left(-\frac{r^2}{r_1^2}\right)rdr \int_0^\infty q_1\exp\left(-\frac{r^2}{r_1^2}\right)2\pi rdr} \quad (9)$$

*For measurement with 20X objective:*

$$R_{m(20X)} = \frac{\int_0^\infty \left(-I_0(z)\lim_{z\to\infty}\frac{\theta_p(z)}{I_0(z)} + \theta_p(z)\right)\exp\left(-\frac{r^2}{r_2^2}\right)rdr}{\int_0^\infty \exp\left(-\frac{r^2}{r_2^2}\right)rdr \int_0^\infty q_2\exp\left(-\frac{r^2}{r_2^2}\right)2\pi rdr} \quad (10)$$

To avoid the artificial shift in mode frequency we have used $R_m = \frac{\partial \theta_m}{\partial Q}$ instead of $R_m \equiv \frac{\theta_m}{Q}$.

And, $R_m$ depends on $\kappa$ and $g$. At a same time, $R_m$ can be experimentally obtained using the following relation

$$R_m = \frac{\partial \theta_m}{\partial Q} = \frac{\partial \omega}{\partial Q}\frac{\partial \theta_m}{\partial \omega} = \chi_P(\chi_T)^{-1} \quad (11)$$

Where $\chi_P$ and $\chi_T$ are first order power and temperature coefficients, respectively.

The $R_m$ values obtained experimentally using equation (11) from the temperature-dependent and power-dependent Raman data for the 50X and 20X objective lenses are used to avoid any calibration error in the κ and g values for the samples by using equation 8. To solve it numerical we have used $R_m = \frac{\partial \theta_m}{\partial Q}$ ratios for 50X and 20X to get accurate results.

**Calculations:**

(i) **For Gr (graphene on SiO$_2$/Si):**

*In case of 50X objective*

$\chi_P$ = -0.993 cm$^{-1}$/mW

$\chi_T$ = -0.035 cm$^{-1}$/K

Therefore, $R_m = \frac{\partial \theta_m}{\partial Q} = \frac{\partial \omega}{\partial Q}\frac{\partial \theta_m}{\partial \omega} = \chi_P(\chi_T)^{-1}$ = 28.371 K/mW

*In case of 20X objective*

$\chi_P$ = -0.308 cm$^{-1}$/mW

$\chi_T$ = -0.009 cm$^{-1}$/K

Therefore, $R_m = \frac{\partial \theta_m}{\partial Q} = \frac{\partial \omega}{\partial Q}\frac{\partial \theta_m}{\partial \omega} = \chi_P(\chi_T)^{-1}$ = 34.222 K/mW

Hence, $R_m$ ratio (of 50X to 20X) = (28.371/34.222) = 0.829

The peak absorbed laser power per unit area (q$_1$) corresponding to 50X objective: $q_1 = \frac{Q_1 \alpha}{\pi r_1^2}$

where,

$Q_1$ = Total absorbed laser power

$r_1$ = Laser spot radius for 50X objective = 0.64 μm

α = Optical absorption coefficient of monolayer graphene = 0.023 [9,12]

$q_1 = \frac{1.38 \times 10^{-3} \times 0.023}{\pi \times (0.64 \times 10^{-6})^2}$ Wm$^{-2}$ = 2.46 × 10$^7$ Wm$^{-2}$

Similarly, for the 20X objective the peak absorbed laser power per unit area (q$_2$): $q_2 = \frac{Q_2 \alpha}{\pi r_2^2}$

where,

$Q_2$ = Total absorbed laser power

$r_2$ = Laser spot radius for 20X objective = 0.94 μm

α = Optical absorption coefficient = 0.023

$q_2 = \frac{2.41 \times 10^{-3} \times 0.023}{\pi \times (0.94 \times 10^{-6})^2}$ Wm$^{-2}$ = $1.99 \times 10^7$ Wm$^{-2}$

Thickness ($l$) of graphene (monolayer) = 0.335 nm.

We have used the information of absorbed power or volumetric heat coefficient ($q_{1,2}$), laser spot radius ($r_{1,2}$), and thickness of the sample ($l$) in the equation (8) with the respective $R_m$ value of the corresponding objective (50X or 20X) and solved the equation (8) numerically to estimate $k$ and g by taking the ratio of $R_m$ for 50X to 20X. The estimated values for Gr (graphene on SiO$_2$/Si) sample are κ = 600.0 Wm$^{-1}$K$^{-1}$ and g = 1.15 MWm$^{-2}$K$^{-1}$

### (ii) **For BN (h-BN on SiO$_2$/Si):**

*In case of 50X objective*

$\chi_P$ = -0.052 cm$^{-1}$/mW

$\chi_T$ = -0.041 cm$^{-1}$/K

Therefore, $R_m = \frac{\partial \theta_m}{\partial Q} = \frac{\partial \omega}{\partial Q} \frac{\partial \theta_m}{\partial \omega} = \chi_P (\chi_T)^{-1}$ = 1.268 K/mW

*In case of 20X objective*

$\chi_P$ = -0.025 cm$^{-1}$/mW

$\chi_T$ = -0.034 cm$^{-1}$/K

Therefore, $R_m = \frac{\partial \theta_m}{\partial Q} = \frac{\partial \omega}{\partial Q} \frac{\partial \theta_m}{\partial \omega} = \chi_P (\chi_T)^{-1}$ = 0.735 K/mW

Hence, $R_m$ ratio (of 50X to 20X) = (1.268/0.735) = 1.725

Now, the peak absorbed laser power per unit area ($q_1$) corresponding to for 50 X objective: $q_1 = \frac{Q_1 \alpha}{\pi r_1^2}$

where,

$Q_1$ = Total absorbed laser power

$r_1$ = Laser spot radius for 50X objective

α = Optical absorption coefficient = 0.015×10= 0.15, [12-14]

$q_1 = \frac{1.38 \times 10^{-3} \times 0.15}{\pi \times (0.64 \times 10^{-6})^2}$ Wm$^{-2}$ = $16.08 \times 10^7$ Wm$^{-2}$

Similarly, for 20 X objective the peak absorbed laser power per unit area ($q_2$): $q_2 = \frac{Q_2 \alpha}{\pi r_2^2}$

where,

$Q_2$ = Total absorbed laser power

$r_2$ = Laser spot radius for 20X objective

α = Optical absorption coefficient

$q_2 = \frac{2.41 \times 10^{-3} \times 0.15}{\pi \times (0.94 \times 10^{-6})^2}$ Wm$^{-2}$ = 13.02 × 10$^7$ Wm$^{-2}$

Thickness ($l$) of 10-layer h-BN = 3.4 nm from AFM data discussed in section S1 above.

We have used the above given information of absorbed power or volumetric heat coefficient ($q_{1,2}$), laser spot radius ($r_{1,2}$), and thickness of the sample ($l$) in the equation (8) with the respective $R_m$ value of the corresponding objective (50X or 20X) and solved the equation (8) numerically to estimate $k$ and g by taking the ratio of $R_m$ for 50X to 20X. The estimated values for BN (h-BN on SiO$_2$/Si) sample are κ = 280.0 Wm$^{-1}$K$^{-1}$ and g = 25.6 MWm$^{-2}$K$^{-1}$

### (iii) **For BG heterostructure (h-BN on graphene on SiO$_2$/Si):**

*In case of 50X objective*

$\chi_P$ = -0.226 cm$^{-1}$/mW

$\chi_T$ = -0.022 cm$^{-1}$/K

Therefore, $R_m = \frac{\partial \theta_m}{\partial Q} = \frac{\partial \omega}{\partial Q} \frac{\partial \theta_m}{\partial \omega} = \chi_P (\chi_T)^{-1}$ = 10.272 K/mW

*In case of 20X objective*

$\chi_P$ = -0.373 cm$^{-1}$/mW

$\chi_T$ = -0.034 cm$^{-1}$/K

Therefore, $R_m = \frac{\partial \theta_m}{\partial Q} = \frac{\partial \omega}{\partial Q} \frac{\partial \theta_m}{\partial \omega} = \chi_P (\chi_T)^{-1}$ = 10.970 K/mW

Hence, $R_m$ ratio (of 50X to 20X) = (10.272/10.970) = 0.936

Now, for 50X the peak absorbed laser power per unit area (q$_1$): $q_1 = \frac{Q_1 \alpha}{\pi r_1^2}$

where,

$Q_1$ = Total absorbed laser power

$r_1$ = Laser spot radius for 50X objective

α = Optical absorption coefficient = 0.15 + 0.023 = 0.173 (It is assumed that the net absorption coefficient of the BG heterostructure will be reasonably close to the sum of the two. To note that small differences in the values of absorption coefficient do not significantly affect the overall estimation of the $k$ and $g$. Hence, our assumption is reasonably acceptable.) [9,12-14]

$$q_1 = \frac{1.38 \times 10^{-3} \times 0.173}{\pi \times (0.64 \times 10^{-6})^2} \text{ Wm}^{-2} = 18.550 \times 10^7 \text{ Wm}^{-2}$$

Similarly, for 20 X the peak absorbed laser power per unit area ($q_2$): $q_2 = \frac{Q_2 \alpha}{\pi r_2^2}$

where,

$Q_2$ = Total absorbed laser power

$r_2$ = Laser spot radius for 20X objective

$\alpha$ = Optical absorption coefficient = 0.038

$$q_2 = \frac{2.41 \times 10^{-3} \times 0.173}{\pi \times (0.94 \times 10^{-6})^2} \text{ Wm}^{-2} = 15.019 \times 10^7 \text{ Wm}^{-2}$$

Thickness ($l$) of the BG heterostructure = (3.4+0.335) nm = 3.735 nm (It is approximately considered as the sum of the 11 layers: one monolayer of graphene and 10 layers of h-BN).

We have used the information of absorbed power or volumetric heat coefficient ($q_{1,2}$), laser spot radius ($r_{1,2}$), and thickness of the sample ($l$), as given above, in the equation (8) with the respective $R_m$ value of the corresponding objective (50X or 20X) and solved the equation (8) numerically to estimate $k$ and $g$ by taking the ratio of $R_m$ for 50X to 20X. The estimated values for BG heterostructure (h-BN on graphene on SiO$_2$/Si) sample are $\kappa$ = 1072 Wm$^{-1}$K$^{-1}$ and g = 26.2 MWm$^{-2}$K$^{-1}$

(iv) **For BGB heterostructure (h-BN on graphene on h-BN on SiO$_2$/Si):**

*In case of 50X objective*

$\chi_P$ = -0.671 cm$^{-1}$/mW

$\chi_T$ = -0.098 cm$^{-1}$/K

Therefore, $R_m = \frac{\partial \theta_m}{\partial Q} = \frac{\partial \omega}{\partial Q} \frac{\partial \theta_m}{\partial \omega} = \chi_P (\chi_T)^{-1}$ = 6.840 K/mW

*In case of 20X objective*

$\chi_P$ = -0.437 cm$^{-1}$/mW

$\chi_T$ = -0.100 cm$^{-1}$/K

Therefore, $R_m = \frac{\partial \theta_m}{\partial Q} = \frac{\partial \omega}{\partial Q} \frac{\partial \theta_m}{\partial \omega} = \chi_P (\chi_T)^{-1}$ = 4.37 K/mW

Hence, $R_m$ ratio (of 50X to 20X) = (6.84/4.37) = 1.565

Now, for 50X the peak absorbed laser power per unit area ($q_1$): $q_1 = \frac{Q_1 \alpha}{\pi r_1^2}$

where,

$Q_1$ = Total absorbed laser power

$r_1$ = Laser spot radius for 50X objective

α = Optical absorption coefficient = 0.15+0.023+0.15 = 0.323 (We have once again assumed that the effective absorption coefficient of the BGB heterostructure will be the sum of the coefficients of the three layers. Further, small differences in the values of absorption coefficient do not significantly affect the overall estimation of the *k* and g. Hence, our assumption is reasonably acceptable.) [9,12-14]

$q_1 = \frac{1.38 \times 10^{-3} \times 0.323}{\pi \times (0.64 \times 10^{-6})^2}$ Wm$^{-2}$ = 34.630 × 10$^7$ Wm$^{-2}$

Similarly, for 20X the peak absorbed laser power per unit area (q$_2$): $q_2 = \frac{Q_2 \alpha}{\pi r_2^2}$

where,

$Q_2$ = Total absorbed laser power

$r_2$ = Laser spot radius for 20X objective

α = Optical absorption coefficient = 0.15+0.023+0.15 = 0.053

$q_2 = \frac{2.41 \times 10^{-3} \times 0.323}{\pi \times (0.94 \times 10^{-6})^2}$ Wm$^{-2}$ = 28.04 × 10$^7$ Wm$^{-2}$

Thickness (*l*) = (3.4+0.335+3.4) nm = 7.135 nm (Sum of the 21 layers, 10 layers of h-BN, one monolayer of graphene, and another 10 layers of h-BN).

We have used the information of absorbed power or volumetric heat coefficient ($q_{1,2}$), laser spot radius ($r_{1,2}$), and thickness of the sample (*l*), as given above, in the equation (8) with the respective R$_m$ value of the corresponding objective (50X or 20X) and solved the equation (8) numerically to estimate *k* and g by taking the ratio of R$_m$ for 50X to 20X. The estimated values for BGB heterostructure (h-BN on graphene on h-BN on SiO$_2$/Si) sample are κ = 850.0 Wm$^{-1}$K$^{-1}$ and g = 105 MWm$^{-2}$K$^{-1}$

### S5. Dependence of surface contact area and surface roughness on introducing h-BN

The schematic given below (Figure S4) depicts how roughness of the graphene layer decreases with the increment of the h-BN layer number. From previous reports [15,16], we know that SiO$_2$/Si has rougher surface than h-BN layer on SiO$_2$/Si because of the presence of the dangling bonds and corrugations in the SiO$_2$/Si substrate, whereas the surface of h-BN layer is free from the draggling bonds [15, 16]. Therefore, the contact area between graphene and the substrate increases, thus increasing the heat dissipation channels through the h-BN layer which increases the interfacial thermal conductance per unit area in BGB by a greater extent.

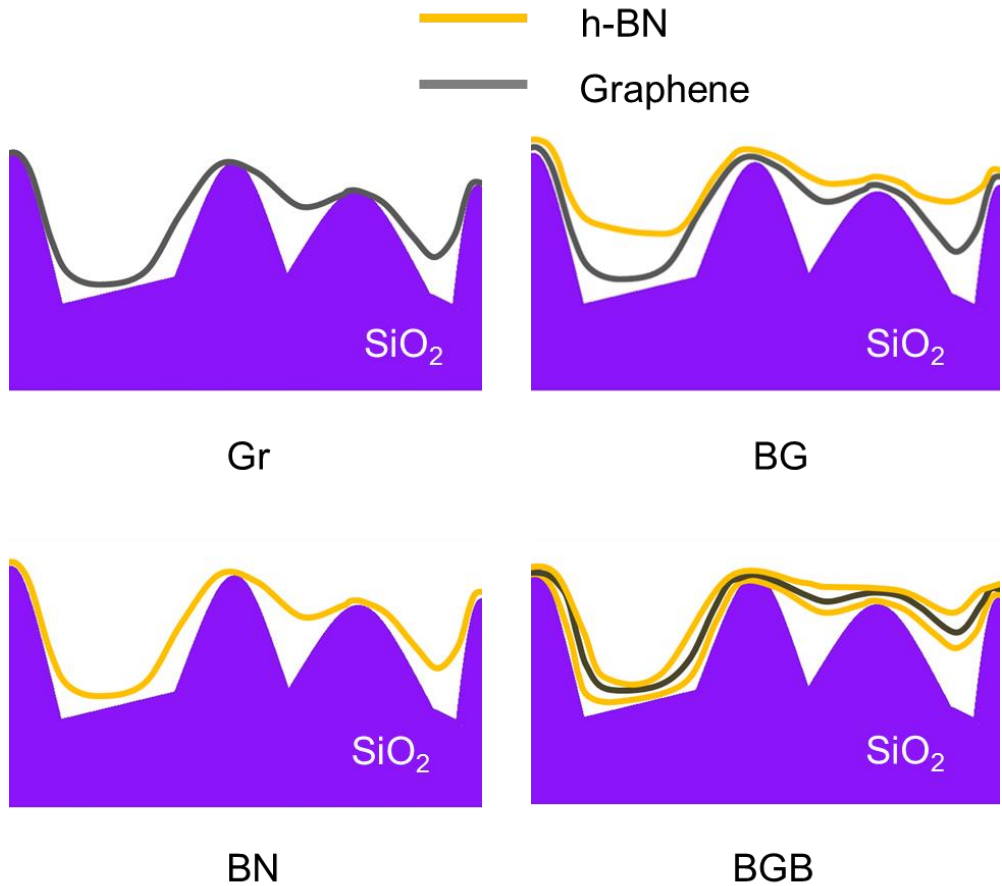

*Figure S4: A schematic showing how the roughness seen by graphene varies as an h-BN layer in introduced in the four (hetero)structures samples.*

### S6. Effect of flake thickness on κ and g

In our work, we have seen an overall improvement in thermal dissipation properties of graphene by using them in conjugation with h-BN as van-der Waals heterostructures. Both the κ (in-plane thermal conductivity) and g (interfacial thermal conductance per unit area) values of the hybrid structures show enhancement. It must be noted that due to the nanoscale thickness of the 2D materials, the dissipation of heat through cross-planar channels can be very strong for supported graphene-based devices. To employ this property of supported devices, it is important to ensure excellent surface contact with the substrate. Thinner flakes are often prone to defects like folding and corrugation, which in turn result in a decreased interfacial thermal conductance. Therefore, it is often preferable to have multilayer flakes in devices to produce better surface contact and higher interfacial thermal conductance. Yuan *et al.* [17] demonstrated an increase in the interfacial thermal conductance in c-Si supported $MoS_2$ with increasing thickness of $MoS_2$ flakes. On the other hand, Jo *et al.* [18] showed an increase in the in-plane thermal conductivity ($\kappa$) of suspended h-BN with increase in the h-BN flake thickness. In our heterostructures, we have therefore, used h-BN layers of thickness ~3.4 nm (10 layers, which may be still considered in the 2D limit) with monolayer graphene.

**S7. Phonon lifetime vs linewidth of Raman modes**

The lifetime of a phonon mode is inversely proportional to the FWHM of the mode ($\Gamma \propto \frac{1}{\tau}$)[19]. Figure S5 shows the FWHMs of the 2D mode of the different samples as a function of temperature. Here we observe that the linewidth corresponding to the 2D mode in Gr, BG, and BGB samples follow the trend:

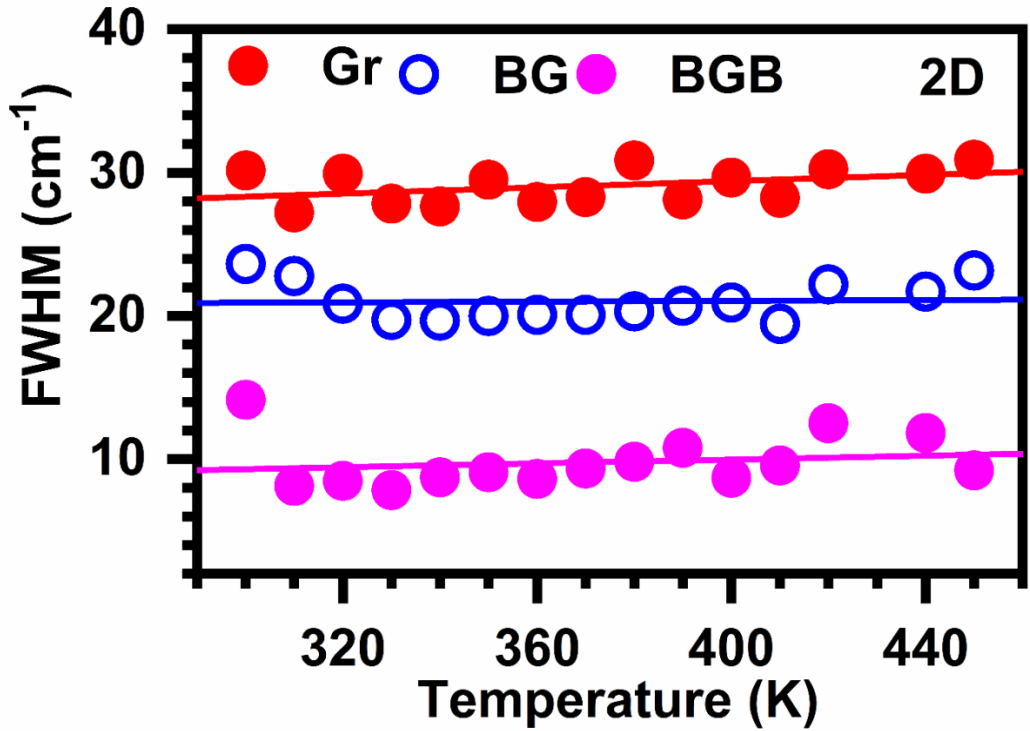

Gr > BG > BGB, which implies that the trend for phonon lifetimes is as BGB > BG > Gr. This is consistent with the increasing trend of phonon lifetimes with increasing number of h-BN and graphene interfaces as seen in Figure 6 of main text.

*Figure S5: FWHM of the 2D mode as a function of temperature for the different samples.*

**Table S2.** Comparative accounts of thermal conductivity and interface thermal conductance values of developed graphene-based (hetero)structures in the present study with respect to various reports on graphene-based thermal studies.

| Structure | Thermal conductivity ($\kappa$) in $WK^{-1}m^{-1}$ | Interface thermal conductance (g) $MWK^{-1}m^{-2}$ | Method | Supported/ Suspended | Reference |
|---|---|---|---|---|---|
| h-BN/SiO$_2$/Si (BN) | 280 | 25.6 | Optothermal Raman method (Experiment) | Supported | Present work |
| Gr/SiO$_2$/Si (Gr) | 600 | 1.15 | | | |
| h-BN/Gr/SiO$_2$/Si (BG) | 1072 | 26.2 | | | |
| Gr/h-BN/Gr/SiO$_2$/Si (BGB) | 850 | 105 | | | |
| Gr/SiO$_2$/Si | 5000 | - | Optothermal Raman method (Experiment) | Suspended | *Nano Letters.* **2008**, *8*, 902 |
| Gr/SiO$_2$/Si | 600 | - | Electrical heating (Experiment) | Supported | *Science* **2010**, *328*, 213 |
| Gr/ SiC | - | 0.02 | Raman thermometry combined with Joule heating (Experiment) | Supported | *Small* **2011**, *7*, 3324 |
| Gr/Au/SiO$_2$/Si | 370 | 28 | Optothermal Raman method (Experiment) | Supported | *Nano Lett.* **2010**, *10*, 1645. |
| Gr/h-BN | 1347.3 | - | Theoretical prediction | - | *Nanotechnology* **2017**, *28*, 225704 |
| Gr/h-BN | - | 6420 | Theoretical prediction on lateral heterostructures | - | *Nano letters.* **2016**, *16*, 4954 |
| Gr/h-BN | - | 7.4 | Raman thermometry combined with Joule | Supported | *Appl. Phys. Lett.* **2014**, *104*, 081908 |

| | | | heating (Experiment) | | |
|---|---|---|---|---|---|